\documentclass[letter]{aa} 
\usepackage[varg]{txfonts}
\usepackage{graphicx}
\usepackage{txfonts}
\usepackage{natbib}
\bibpunct{(}{)}{;}{a}{}{,} 
\begin{document}

\title{Forever young white dwarfs: when stellar ageing stops}

\author{Mar\'ia E. Camisassa\inst{1,2,3},
	Leandro G. Althaus\inst{2,4}, 
	Santiago Torres \inst{3,5},
	Alejandro H. C\'orsico\inst{2,4},
    Alberto Rebassa-Mansergas\inst{3,5},
	Pier-Emmanuel Tremblay \inst{6},
	Sihao Cheng \inst{7}
    \and
    Roberto Raddi \inst{3}} 
\institute{Department of Applied Mathematics, University of Colorado, Boulder, CO 80309-0526, USA
           \and
           Instituto de Astrof\'isica de La Plata, UNLP-CONICET, 
           Paseo  del Bosque s/n, 1900 
           La Plata, 
           Argentina
     	    \and
           Departament de F\'\i sica Aplicada, 
           Universitat Polit\`ecnica de Catalunya, 
           c/Esteve Terrades 5, 
           08860 Castelldefels, 
           Spain
           \and
           Facultad de Ciencias Astron\'omicas y Geof\'{\i}sicas, 
           Universidad Nacional de La Plata, 
           Paseo del Bosque s/n, 1900 
           La Plata, 
           Argentina
           \and
           Institute for Space Studies of Catalonia,  
           c/Gran Capita 2--4, Edif. Nexus 201, 
           08034 Barcelona,  Spain
           \and
           Department of Physics, University of Warwick, Coventry, CV4 7AL, UK
           \and
          Department of Physics and Astronomy, The Johns Hopkins University, 3400 N Charles Street, Baltimore, MD 21218, USA}
\date{Received ; accepted }

\abstract{White dwarf stars are the most common end point of stellar evolution. Of special interest are the ultramassive white dwarfs, as they are related to type Ia Supernovae explosions, merger events, and Fast Radio Bursts. Ultramassive white dwarfs are expected to harbour oxygen-neon (ONe) cores as a result of single standard stellar evolution. However, a fraction of them could have carbon-oxygen (CO) cores. Recent studies, based on the new observations provided by the {\it Gaia} space mission, indicate that a small fraction of the ultramassive white dwarfs experience a strong delay in their cooling, which cannot be attributed only to the occurrence of crystallization, thus requiring an unknown energy source able to prolong their life for long periods of time. In this study we { find} that the energy released by $^{22}$Ne sedimentation in the deep interior of ultramassive white dwarfs with CO cores { and high $^{22}$Ne content is consistent with the long cooling delay of these stellar remnants. On the basis of a synthesis study of the white dwarf population, based on Monte Carlo techniques, we find that the observations revealed by {\it Gaia} can be explained by the existence of these prolonged youth ultramassive white dwarfs. Although such a high $^{22}$Ne abundance is not consistent with the standard evolutionary channels, our results provide sustain} to the existence of CO-core ultramassive white dwarfs and to the occurrence of $^{22}$Ne sedimentation.}
\keywords{stars:  evolution  ---  stars: interiors  ---  stars:  white
  dwarfs}
\titlerunning{Forever young white dwarfs}
\authorrunning{Camisassa et al.}  

\maketitle

\section{Introduction}
\label{introduction}
White dwarfs (WD) are the most common fossil stars within the stellar graveyard \citep{2010A&ARv..18..471A}. It is well known that more than 95\% per cent of all main-sequence stars will finish their lives as WDs, earth-sized objects less massive than $\sim$1.4 M$_\odot$---the Chandrasekhar limiting mass--- supported by electron degeneracy \citep{1931ApJ....74...81C}. A remarkable property of the WD population is its mass distribution, which exhibits a main peak at $\sim$0.6 M$_\odot$, a smaller peak at the tail of the distribution around  ~0.82 M$_\odot$, and a low-mass excess near $\sim$ 0.4 M$_\odot$ \citep{2015MNRAS.452.1637R,2013ApJS..204....5K,2018MNRAS.480.4505J}. WDs with masses lower than 1.05 M$_\odot$ are expected to harbour carbon(C)-oxygen(O) cores, enveloped by a shell of helium which is surrounded by a layer of hydrogen. WDs more massive than 1.05 M$_\odot$ are called ultramassive WDs and, conventionally, they are expected to contain an oxygen-neon(Ne) core. 

The study of ultramassive WDs is motivated by their connection to type Ia Supernova, double WD mergers, the occurrence of physical processes in the Asymptotic Giant Branch, and the existence of high magnetic fields.
Traditionally, the formation of ultramassive WDs is theoretically predicted as the end product of the isolated evolution of intermediate-mass stars with an initial mass larger than 6-9 M$_\odot$  \citep{2007A&A...476..893S,2010A&A...512A..10S}.
However, in the recent years, the literature has been filled out with evidence supporting that a large fraction of the single ultramassive WDs could be the formed through binary evolutionary channels. Recent studies \citep{2017A&A...602A..16T,2018MNRAS.476.2584M} suggest that binary mergers contribute substantially to the single WD population. The theoretical studies of \cite{2020A&A...636A..31T} estimated that 30–45\% of the WDs more  massive  than  0.9M$_\odot$ are  formed  through  binary  mergers,  mostly  via  the  merger of two WDs, and \cite{2020ApJ...891..160C} estimated observationally that the fraction of high-mass WDs (in  the  range  0.8–1.3M$_\odot$) formed as a result of double WD mergers is nearly 20 \%.

The core chemical composition of an ultramassive WD is still a matter of debate.
Traditionally, in the single evolution channel, once the helium in the core has been exhausted, massive intermediate-mass stars evolve to the super asymptotic giant branch (SAGB) phase, where they reach temperatures high enough to start off-centre carbon ignition under partially degenerate conditions \citep{2017PASA...34...56D}. A violent carbon
ignition eventually leads to the formation of an ONe core, that will ultimately become an ultramassive ONe WD. Nevertheless, \cite{2021A&A...646A..30A} have shown that rotation or reduced mass loss rate during the SAGB would prevent the carbon ignition, leading to the formation of ultramassive CO core WDs from single evolution.
 For the case of WDs that result from double WD mergers, \cite{2007MNRAS.380..933Y} and \cite{2009A&A...500.1193L}, predict that the merger remnant avoids carbon burning and becomes a single CO-core ultramassive WD. In the opposite direction, the recent study of \cite{2020arXiv201103546S}, based on one-dimensional post merger evolutionary calculations, predicts the off-centre carbon burning in the merged remnant, leading to the formation of an ONe-core ultramassive WD.
To date, it has not been possible to distinguish a CO-core from a ONe-core ultramassive WD from their observed properties, although a promissory avenue to accomplish this is by means of WD asteroseismology \citep{2019A&ARv..27....7C}. 

	\begin{figure}
	\centering
	\includegraphics[trim=40 45 0 0,clip,width=\columnwidth]{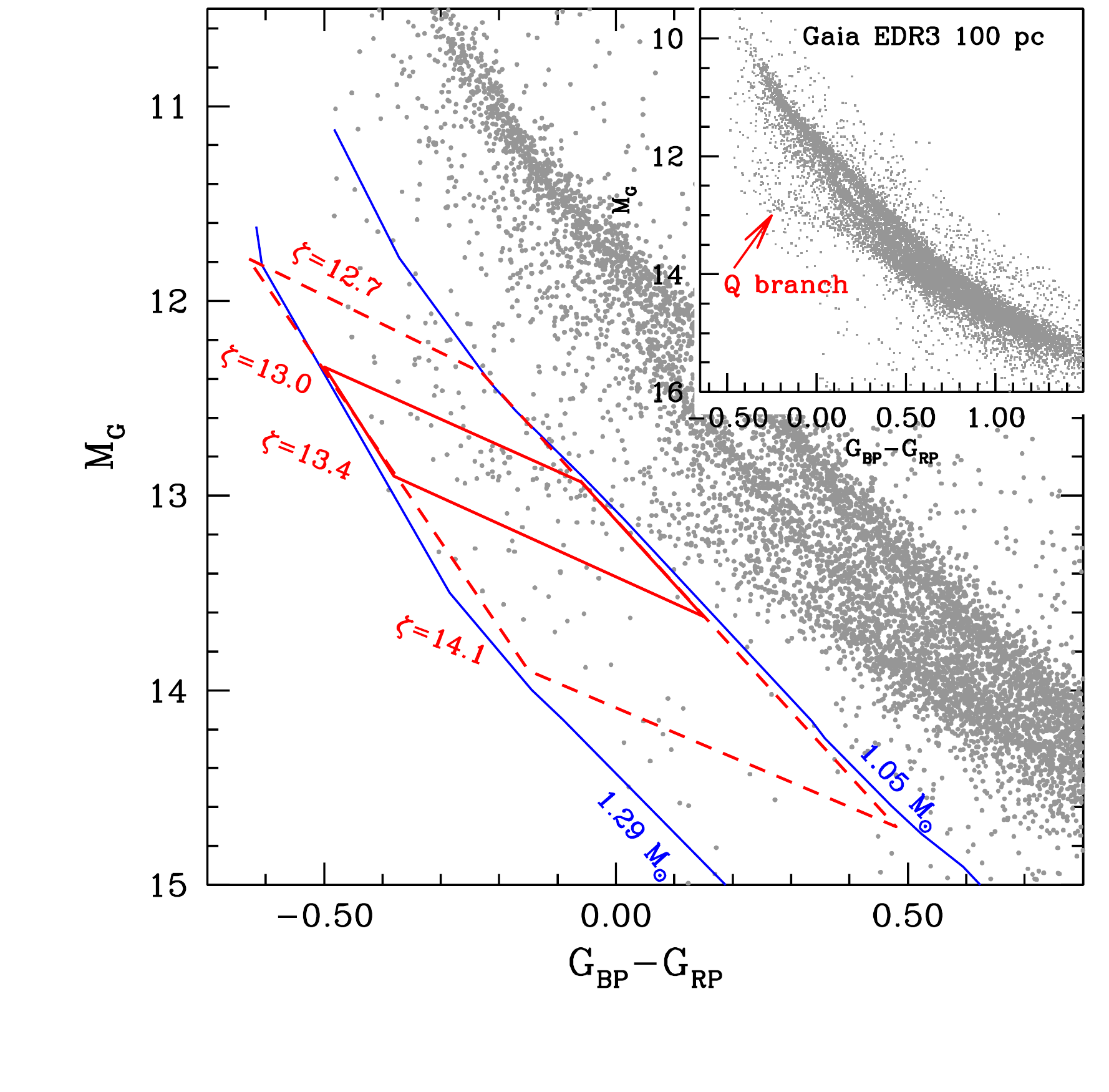}\\
    \caption{
    The WD sequence (grey points) in the {\it Gaia} Hertzsprung-Russell diagram. Theoretical ONe-core WD cooling tracks of 1.05 and 1.29 M$_\odot$ disregarding $^{22}$Ne diffusion are shown as solid blue lines \citep{2019A&A...625A..87C}. The Q branch is represented as an observed overabundance of WDs below the standard cooling sequence. In order to mark off the Q branch, we have defined the parameter ${\rm \zeta=M_G-1.2\times(G_{BP}-G_{RP})}$. The ultramassive Q branch is delimited by $\zeta=13.4$ and $\zeta=13.0$, and by the cooling tracks of 1.05 and 1.29 M$_\odot$. Dashed red lines delimit the region where we have counted WDs to prepare the histograms.}
    \label{fig:1}
\end{figure}
	
Since April 2018, WDs have gathered a renewed interest in the scientific community, as the Hertzprung-Russell (HR) diagram provided by the {\it Gaia} Data Release 2 \citep{2018A&A...616A..10G} has revealed some unexpected features, that have been later reconfirmed by the {\it Gaia} Early Data Release 3 (EDR3) \citep{GaiaDR3}. One of them corresponds to a { transverse branch} in the WD cooling sequence, called the Q branch, that can be appreciated by inspecting Figure \ref{fig:1}. Defining the parameter $\zeta=\rm M_G-1.2\times(G_{BP}-G_{RP})$, which lies parallel to the Q branch, the ultramassive Q branch is delimited by $\zeta$=13.0 and $\zeta$=13.4, and by the cooling tracks of 1.05 and 1.29 M$_\odot$. 
Recently, the Q branch has been attributed to the crystallization process occurring in the interior of WDs due to Coulomb interactions \citep{2019Natur.565..202T}. The crystallization process in a WD not only releases energy as latent heat, but also releases gravitational energy due to a phase separation process that alters the stellar chemical abundances in the core. These two energy sources delay the cooling of WDs for long periods of time. However, crystallization is expected to be a less relevant process in the evolution of ultramassive WDs \citep{2019A&A...625A..87C}. In fact, a more recent study of the ultramassive Q branch that considers the age velocity dispersion relation shows that models taking into account both the energy released by latent heat and phase separation due to crystallization fail in accounting for the pile-up of ultramassive WDs on the Q branch \citep{2019ApJ...886..100C}. In particular, it is estimated that a fraction of the ultramassive WD population should experience an unusual delay of ~8 Gyrs in their cooling times during their stay on the Q branch, when compared with evolutionary models that consider the energy released by the crystallization process \citep{2019A&A...625A..87C}. This implies the need of an additional energy source, able to produce strong delays in the WD cooling times.  \cite{2020PhRvD.102h3031H} tried unsuccessfully to explain these time delays as a result of electron capture and pycnonuclear reactions and dark matter heating. On the other hand, \cite{2020ApJ...902...93B} claimed that these strong delays could be the result of clustering-enhanced gravitational sedimentation. 
However, the recent paper of \cite{2020ApJ...902L..44C} questions these results using molecular-dynamics simulations, demanding new answers to this problem.
	
	In this paper, we study the sedimentation of $^{22}$Ne occurring in the interior of ultramassive WDs both with CO and ONe composition. We have performed a population synthesis study and we found that these strong delays in the cooling times reported for a selected population of the ultramassive WDs { can be explained as a consequence of} the energy released by the sedimentation process of $^{22}$Ne occurring in the interior of CO-core ultramassive WDs with high $^{22}$Ne content. 
	
\section{Methods}
\label{Methods}

We have calculated a complete set of ultramassive WD models for both CO and ONe core composition of 1.10, 1.16, 1.22 and 1.29 M$_\odot$, varying  the $^{22}$Ne mass fractions from ${\rm X_{^{22}Ne}=0.001}$ to ${\rm X_{^{22}Ne}=0.06}$. The evolutionary calculations presented in this paper were done with an updated version of the LPCODE stellar evolutionary code (see \cite{2005A&A...435..631A} and references therein). This code has been well tested and calibrated with other stellar evolutionary codes in different evolutionary phases, such as the red giant phase \citep{2020A&A...635A.164S,2020A&A...635A.165C} and
the WD cooling phase \citep{2013A&A...555A..96S}, and has been amply used in the study of different aspects of low-mass star evolution \citep{2017ApJ...839...11C,2016A&A...588A..25M}.


 The  energy  contribution  resulting  from  the  gravitational settling  of $^{22}$Ne  is  treated  in  a  similar  way  as it was  done  in \cite{2016ApJ...823..158C,2010ApJ...719..612A}, assuming that the liquid behaves as a single-background one-component plasma plus traces of $^{22}$Ne. The background matter consists of a fictitious element in which the atomic mass (A) and the atomic charge (Ze) are defined by the average A and Ze in each layer. The changes in the $^{22}$Ne chemical profile and the associated local contribution to the luminosity equation are provided by an  accurate  treatment  of  time-dependent $^{22}$Ne diffusion; see \cite{2008ApJ...677..473G} for details. In particular, in the liquid interior we have considered the diffusion coefficients from \cite{2010PhRvE..82f6401H}, which are the result of molecular dynamic simulations. For  those  regions  of  the  WD models  that  are crystallized, the viscosity is expected to abruptly increase, preventing $^{22}$Ne sedimentation process to operate in these solid regions. Therefore, we have set the diffusion coefficient $D=0$ in the crystallized regions.

We have also considered the energy sources resulting from the crystallization of the  WD core, i.e., the release of latent heat and the release of gravitational energy associated with a phase separation process induced by crystallization. In LPCODE, these energy sources are included self-consistently and are locally coupled to the full set of equations of stellar evolution (see \cite{2019A&A...625A..87C,2010ApJ...719..612A}). The crystallization temperature and the changes in the chemical abundances due to crystallization in CO-core WDs are taken from the phase diagram of \cite{2010PhRvL.104w1101H}, and from the phase diagram of \cite{2010PhRvE..81c6107M} for ONe-core WDs.

For the ONe ultramassive WD models, we have considered the initial chemical profiles that result from the full computation of previous evolutionary stages, calculated in \cite{2007A&A...476..893S,2010A&A...512A..10S}.  For the ultramassive CO-core WDs we have considered the chemical profiles of \cite{2021A&A...646A..30A}. The initial $^{22}$Ne chemical abundance of our WD models has been artificially set at the beginning of the WD cooling sequence. 

The observed sample consists in 11,707 WDs belonging to the thin disk population. The sample has been selected from  {\it Gaia} EDR3,  and later classified in its Galactic components by means of artificial intelligent techniques based on an accurate Random Forest algorithm (see \cite{2019MNRAS.485.5573T} for the classification technique,
whereas an update to EDR3 sample will be published
in a forthcoming paper). Only objects with accurate astrometric and photometric measurements (relative error lower than 10\%)  and within 100 pc from the Sun were considered. The complete set of selecting criteria can be consulted in \cite{2019MNRAS.485.5573T,2018MNRAS.480.4505J}.

 Our population synthesis code based on Monte Carlo techniques has been widely used in the study of the WD population during the last decades. A comprehensive explanation of the ingredients can be found in \cite{2019MNRAS.485.5573T,2018MNRAS.480.4505J} and references therein. Here we just mention the major physical inputs. Our modelling of the thin disk population consists in a constant star formation history which lasts 9.2 Gyr. Main-sequence stars are randomly drawn according to a Salpeter law. For each star a metallicity value is adopted following a metallicity dispersion law centred at solar metallicity value Z$_\odot$=0.014. Once the mass and the metallicity of each star is known, we derive its main-sequence lifetime from the models of \cite{bastinew} and the WD mass using the initial-to-final mass relationship of \cite{2008MNRAS.387.1693C}. It is possible then to compute which stars have become WDs and its corresponding cooling age.  For the ultramassive WDs we set the percentage of CO-core and ONe-core of the synthetic population, and randomly adopt a chemical composition for each artificial WD. We also set different $^{22}$Ne abundances to model the effect of  $^{22}$Ne sedimentation on the {\it Gaia} 100 pc WD population. Once the mass, the cooling time, the core chemical composition and the $^{22}$Ne abundance are known, the magnitudes in  {\it Gaia} filters are
derived by interpolating the evolutionary models with
the grid of synthetic WD atmospheres detailed by \cite{2010MmSAI..81..921K}, including a number of
more recent unpublished improvements. Finally, observational uncertainties are added to each of our simulated object by introducing photometric and astrometric errors in concordance with {\it Gaia} performances (http://www.cosmos.esa.int/web/gaia/science-performance ).
	
\section{Results}
\subsection{ The cooling age delay caused by 22Ne settling}
	The neutron excess of $^{22}$Ne, relative to the matter composing the WD core that has equal number of protons and neutrons, results in a net downward gravitational force and a slow settling of $^{22}$Ne in the liquid regions towards the centre of the WD \citep{2002ApJ...580.1077D,1991A&A...241L..29I}. $^{22}$Ne sedimentation process releases substantial energy to appreciably modify the cooling of WDs, causing delays in their evolution \citep{2016ApJ...823..158C}. The occurrence of this process in average-mass WDs has been shown to be a key factor in solving the long-standing age discrepancy of the metal-rich cluster NGC 6791 \citep{2010Natur.465..194G}.
	
		\begin{figure}
		\centering
	\includegraphics[clip,width=\columnwidth]{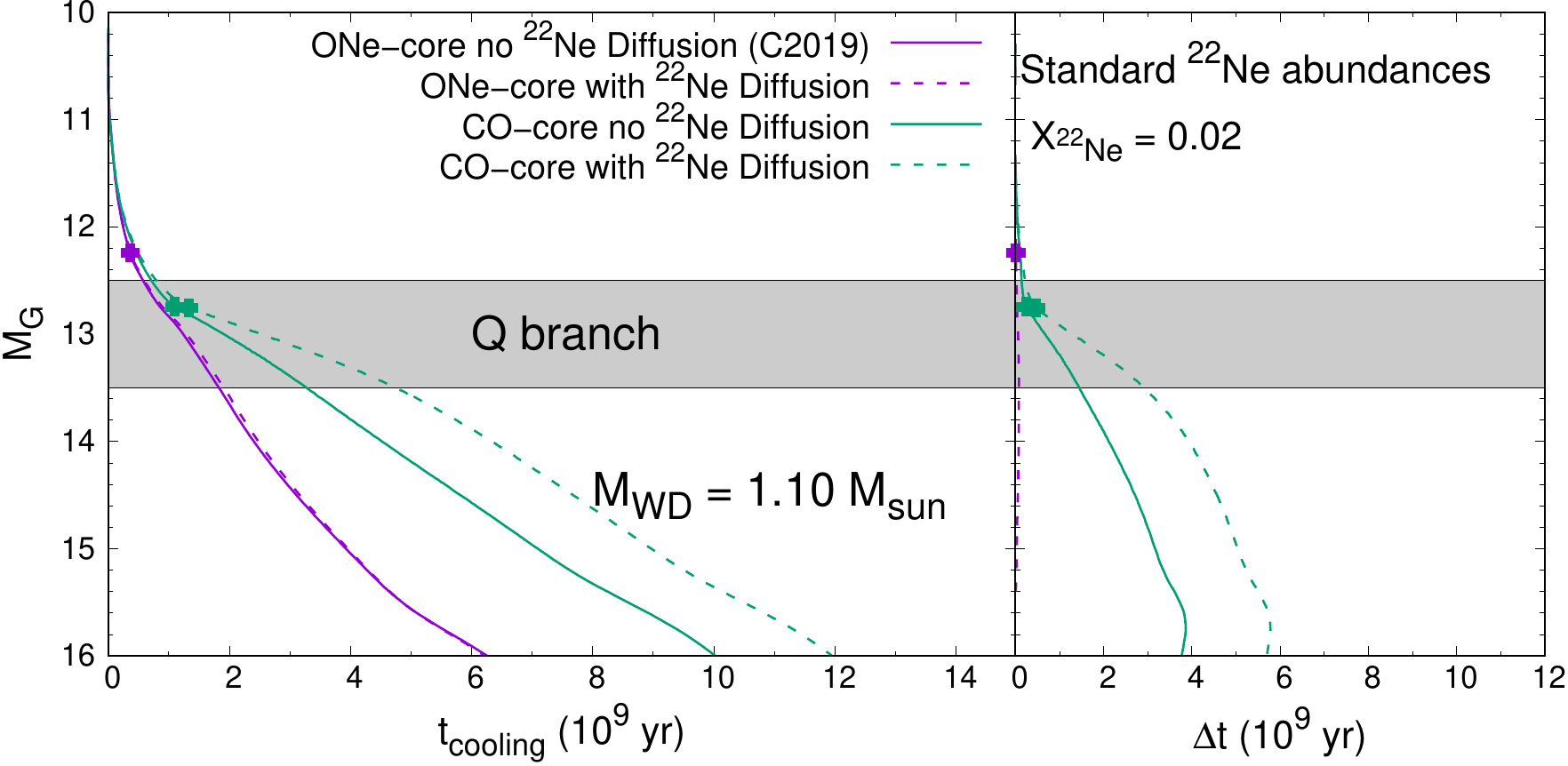}
    \caption{Impact of $^{22}$Ne diffusion on the WD cooling times, defined as the time since the star reaches its maximum effective temperature at the beginning of the WD phase, for cooling sequences with standard $^{22}$Ne abundances and different core chemical compositions. Solid lines indicate the ONe- and CO-core cooling sequences of 1.10 M$_\odot$ disregarding $^{22}$Ne diffusion, whilst dashed lines illustrate the behaviour for 1.10 M$_\odot$ sequences with ONe and CO cores that consider the impact of $^{22}$Ne diffusion. The right panel displays the resulting time delays (in Gyrs) relative to the 1.10 M$_\odot$ ONe-core WD cooling track of \cite{2019A&A...625A..87C} that disregards $^{22}$Ne diffusion process.  Filled squares indicate the onset of core crystallization in each WD sequence and the shaded area indicates when the model is overpassing the Q branch, i.e., the region where the cooling delays are revealed by the observations of {\it Gaia}. 
    }
    	    \label{fig:2}
\end{figure}
	
	The diffusion coefficient of $^{22}$Ne in the liquid core of WDs has a strong dependence on the atomic charge Ze of the background matter \citep{2001ApJ...549L.219B,2010PhRvE..82f6401H}. Therefore, $^{22}$Ne sedimentation turns out to be more efficient in a CO-core WD than in a ONe-core WD, thus resulting in a selective effect that produces strong delays in the cooling times of CO-core WDs, but not in ONe-core WDs. In Figure \ref{fig:2}, we show the effect of $^{22}$Ne diffusion process on the evolution of selected 1.10 M$_\odot$ ultramassive WD sequences with different core chemical compositions, i. e., CO-core and ONe-core WDs. All these WD sequences consider the energy released by latent heat and phase separation due to the crystallization process. The WD sequences displayed with dashed lines consider also the energy released by $^{22}$Ne sedimentation process. The strong dependence of the diffusion coefficients on the core chemical composition results evident by inspecting this Figure. For the CO-core WD considering $^{22}$Ne diffusion, the time delay compared to the ONe-core WD sequence that disregards $^{22}$Ne sedimentation is on average ~5 Gyr. For the ONe-core WD sequence, the time delay caused by $^{22}$Ne sedimentation process is negligible. Although ultramassive WDs are affected by $^{22}$Ne sedimentation quite early in their evolution due to their large characteristic gravities, this process is abruptly interrupted in the crystallized core, and the energy released is reduced when crystallization process sets in. Coulomb interactions are weaker in nuclei with lower atomic charge, and thus, crystallization occurs at lower luminosities in CO-core WDs than in their ONe-core counterparts, allowing $^{22}$Ne diffusion to operate for a longer period of time.
	{ We have performed artificial calculations in order to test if the longer cooling delays
	obtained for CO WDs are mainly due to the fact that these WDs crystallize at lower luminosities or
	due to the fact that the diffusion coefficient in these stars is larger, and we found that this last factor was crucial.} 
	
	On the other hand, the total thermal content of a WD is higher for lower atomic-mass chemical compositions. Hence, the thermal energy stored in the ions is higher in CO-core WDs than in ONe-core WDs, reducing their cooling rate. This fact is reflected in the long cooling times experienced by the CO-core WD model that does not consider $^{22}$Ne sedimentation process in Figure \ref{fig:2}.  However, the delays induced only by considering CO-core chemical composition and not including $^{22}$Ne diffusion are not long enough to account for the observations revealed by the {\it Gaia} space mission.

		\begin{figure}
		\centering
	\includegraphics[clip,width=\columnwidth]{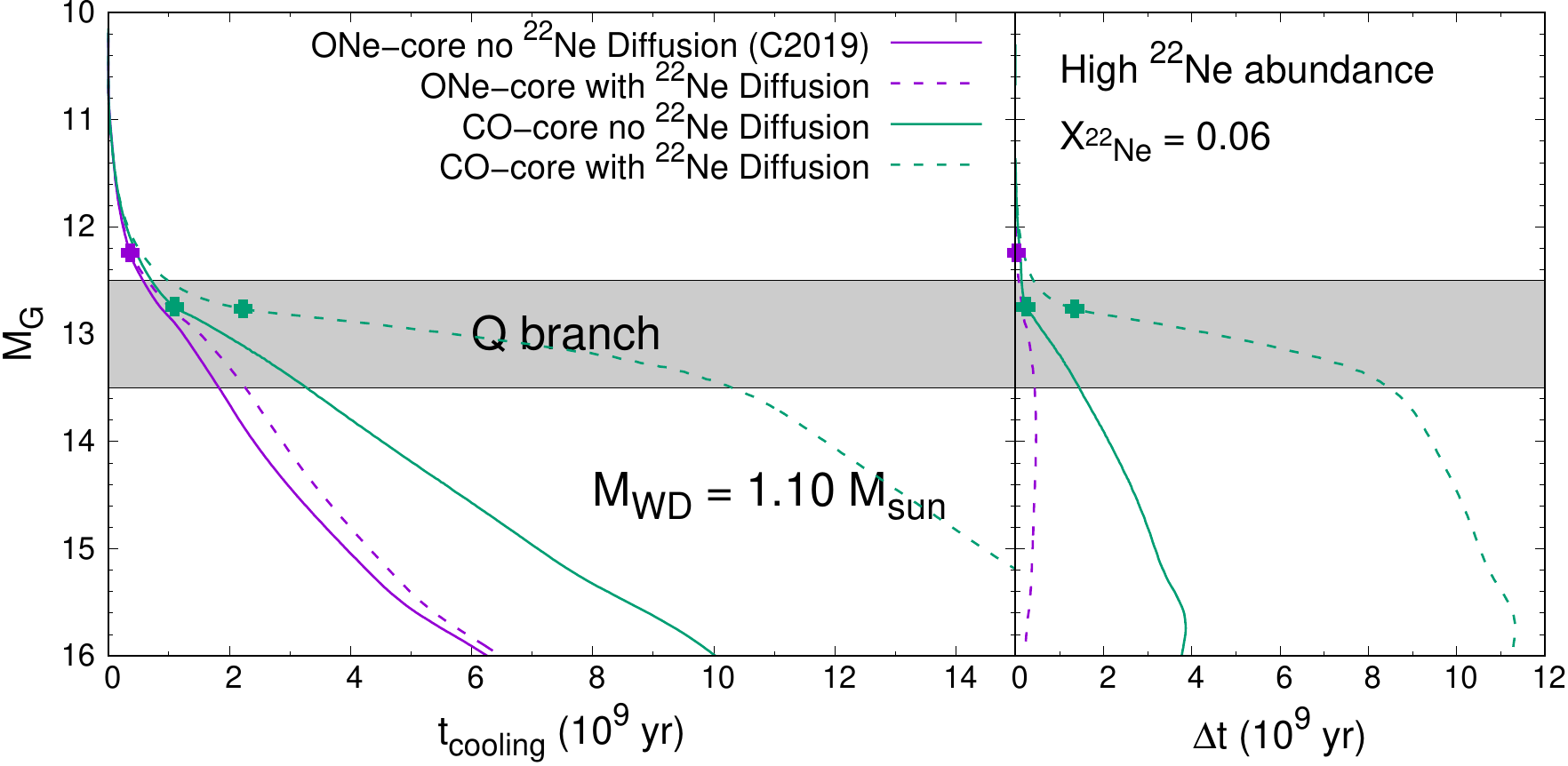}
    \caption{Same as Figure \ref{fig:2}, but for ultramassive WD models with high $^{22}$Ne initial abundances, i.e., ${\rm X_{^{22}Ne}=0.06}$. 
    }
   \label{fig:3}
\end{figure}

	
	To correctly assess the WD cooling delays induced by $^{22}$Ne sedimentation, the amount of $^{22}$Ne content in these stars is crucial, as higher $^{22}$Ne abundances will lead to larger delays in the cooling times  \citep{2010Natur.465..194G}. In the single-star evolution scenario, the abundance of $^{22}$Ne expected in a WD star is roughly equal to the initial stellar metallicity, and is created during the helium core burning phase from helium captures on $^{14}$N, via the reactions $\rm{^{14}N(\alpha,\gamma)^{18}F(\beta+)^{18}O(\alpha,\gamma)^{22}Ne}$. $^{14}$N is left from hydrogen burning via the CNO cycle. Hence, $^{22}$Ne sedimentation will be more effective in those WDs that result from high-metallicity progenitors , due to their larger $^{22}$Ne content. On the contrary, the amount of $^{22}$Ne remaining in the core of a WD that results of a merger event could be larger, depending on the hydrogen content of the binary components at the moment of merger. The study of \cite{2012ApJ...757...76S}  shows that the abundance in a merger episode of two WDs could be as high as ~0.06. { In order to maximize the effect of $^{22}$Ne sedimentation}, we have also explored { this process} considering a higher $^{22}$Ne abundance (${\rm X_{^{22}Ne}}=0.06$), and we confirmed the strong dependence of the cooling delays of CO-core ultramassive WDs on the $^{22}$Ne content.  The impact of $^{22}$Ne sedimentation process on the evolution of CO core ultramassive WDs is particularly strong for high $^{22}$Ne content, where the delays in the cooling times reach 8 Gyrs at the Q branch, and 11 Gyrs by the faint end of the cooling sequence, when compared with the ONe-core cooling sequence (see Figure \ref{fig:3}). On the other hand, due to the larger mean atomic number in their core chemical composition, the cooling of ONe core ultramassive WDs is not substantially altered by $^{22}$Ne sedimentation process, regardless of their $^{22}$Ne content. The cooling age of a 1.10M$_\odot$ ultramassive ONe WD in the Q branch is small, whereas a CO-core WD with ${\rm X_{^{22}Ne}=0.06}$ of the same mass amounts to 10 Gyrs (see Figure \ref{fig:3}). These strong delays in the cooling times of CO-core ultramassive WDs imply that these stars will remain in the Q branch for longer periods of time, when compared with their ONe counterparts, that will rapidly age towards fainter absolute magnitudes.  
	
	\subsection{Population synthesis analysis}
\begin{figure}
		\centering
	\includegraphics[clip,width=\columnwidth]{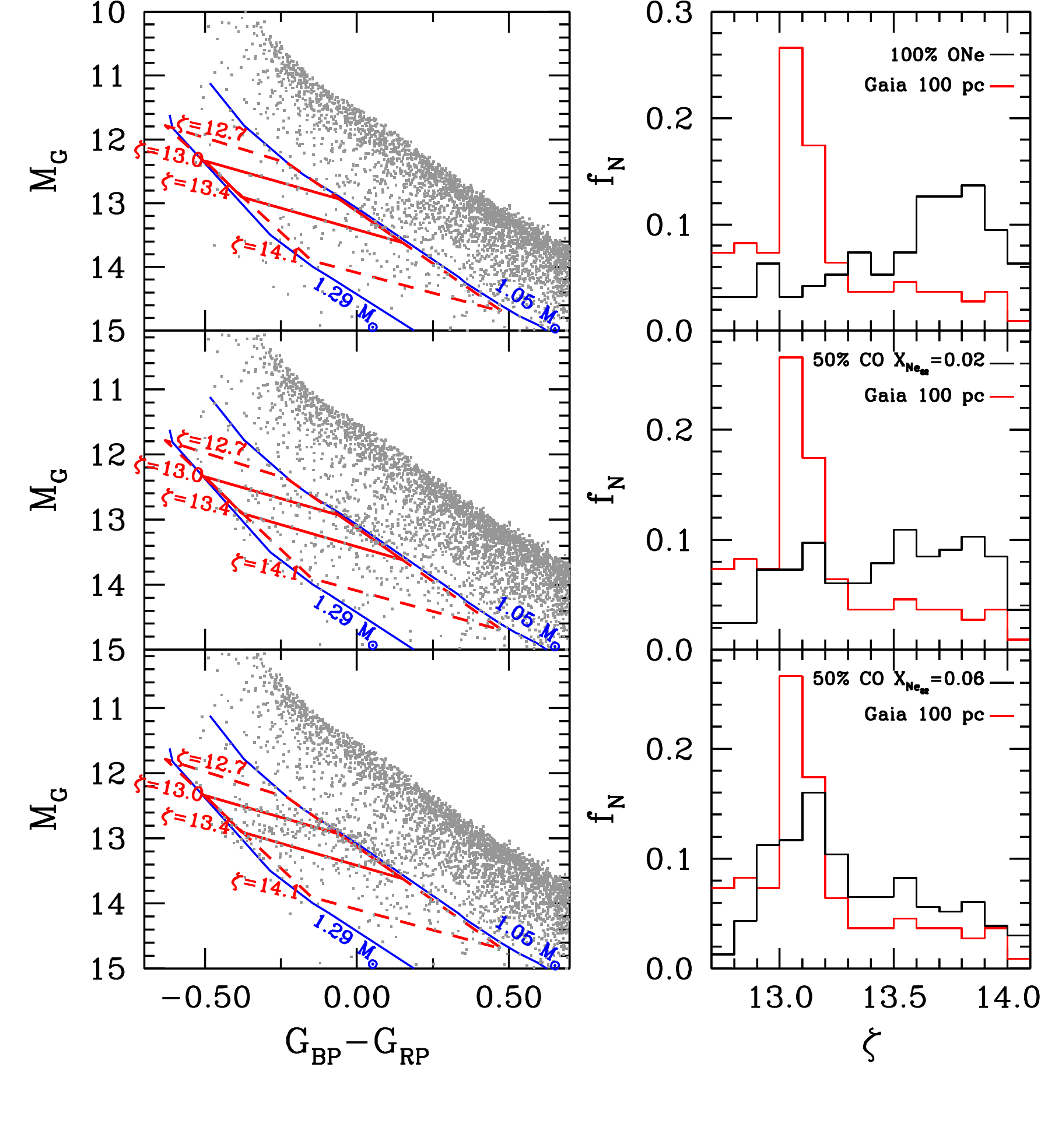}
    \caption{Left panels: Synthetic WD populations (grey points) in the {\it Gaia} Hertzsprung-Russell diagram considering different prescriptions. The ultramassive Q branch is delimited by solid red lines. Dashed red lines mark the region where we have counted WDs to prepare the histograms. Right panels: synthetic (black line) and observed (red line) histograms for the 100pc WD population. The observed Q branch can be regarded as the red peak between $\zeta=13.0$ and $\zeta=13.4$ in the 100 pc WD sample observed by {\it Gaia}.
    }
    	    \label{fig:4}
\end{figure}

	In order to demonstrate the possible existence of these eternal youth ultramassive CO-core WDs, we have analysed the effect of $^{22}$Ne sedimentation on the local
 {\it Gaia}-selected WD population of the Galactic thin disk, within 100 pc from the Sun, by
means of an up-to-date population synthesis code. Accounting for a range of input conditions, we
first considered that all the ultramassive WDs in the simulated sample have ONe core composition. This first synthetic population is shown in the simulated {\it Gaia} HR diagram in the upper left panel of Figure \ref{fig:4}. The histogram distribution of $\zeta$ of this synthetic population is shown in the upper right panel (black steps), together with the {\it Gaia} 100 pc WD sample (red steps). The Q branch can easily be regarded as the main peak in the histogram of the {\it Gaia} 100 pc WD sample, between $\zeta=13.0$ and $\zeta=13.4$. A first glance at these histograms reveals that ultramassive ONe WDs fail to account for the pile-up in the Q branch, even though the ONe WD sequences used include all the energy sources resulting from the crystallization process. A quantitative statistical reduced $\chi^2$-test analysis of the synthetic population distribution in the Q branch reveals a value of 4.27 when compared to the observed distribution.
	
	The middle left panel of Figure \ref{fig:4} illustrates the HR diagram of a synthetic WD population realization, considering that 50\% of the ultramassive WDs harbour a CO-core. In this model we also have assumed WDs with standard $^{22}$Ne abundance, ${\rm X_{^{22}Ne}}=0.02$. The histogram of this synthetic population
	reveals that, although a mixed WD population with both CO-core and ONe-core WDs is in better agreement with the observations, the pile-up is still not fully reproduced. The reduced $\chi^2$ value of this synthetic population is 3.37. 

	Finally, we have performed another population synthesis realization in which 50\% of the ultramassive WDs have a CO core-chemical composition, but this time considering that the $^{22}$Ne content of the CO-core WD sequences is ${\rm{X_{^{22}Ne}}}=0.06$. The results of this synthetic population are shown in the lower panels of Figure \ref{fig:4}. We found that this simulation is in a much better agreement with the observed WD sample, being its reduced $\chi^2$-test value 2.01.  
	
	The better agreement with the observations revealed by {\it Gaia} of the synthetic populations that include CO-core ultramassive WD sequences is in line with the longer cooling times that characterize these stars due to $^{22}$Ne sedimentation process. We have also simulated a synthetic population that considers a fraction of CO-core WDs of 80\% and a $^{22}$Ne abundance of 0.02, finding a better agreement when compared to simulations computed with only ONe-core WDs, but not as good as the agreement we found for a population with high $^{22}$Ne content. We have also generated synthetic populations considering different $^{22}$Ne abundances in the WD models and found that the best fit models are obtained for a high $^{22}$Ne abundance (${\rm X_{^{22}Ne}}=0.06$). { We would like to remark that such} a high $^{22}$Ne abundance is not consistent with the isolated standard evolutionary history channel, because it would imply that these  WDs come from high-metallicity progenitors. However, merger events { could} provide a possible scenario to create such a high $^{22}$Ne abundance. If H were burnt in C-rich layers during the merger event, it would create a high amount of $^{14}$N that could later capture He ions, creating a high $^{22}$Ne abundance before the ultramassive WD is born.

	The analysis of the ultramassive WD population revealed by {\it Gaia} shows that ONe-core WDs alone are not able to account for the pile-up in the ultramassive Q branch. Indeed, energy sources as latent heat and phase separation process due to crystallization, and $^{22}$Ne sedimentation can not prevent the fast cooling of these stars. { According to our study, the Q branch might be explained by} the presence of WDs with CO core, that experience a stronger delay in their cooling due to the combination of three effects: crystallization, $^{22}$Ne sedimentation and higher thermal content.

	\section{Conclusions}
	
	In this study, we find that CO-core ultramassive WDs with high $^{22}$Ne content are long-standing living objects, that should stay on the Q branch for long periods of time. Indeed, their CO core composition, combined with a high $^{22}$Ne abundance, provides a favourable scenario for $^{22}$Ne sedimentation to effectively operate, producing strong delays in the cooling times, and leading to an eternal youth source. { Our study indicates that the cooling delays observed by {\it Gaia} space mission provide valuable sustain to the existence of ultramassive WDs with CO chemical composition, whilst ONe core WDs are unable to predict these delays.} This work is the first population synthesis study based on Monte Carlo techniques that uses the luminosities of the ultramassive WDs measured by {\it Gaia}, { and it reveals that {\it Gaia} observations are consistent with a population in which 50\% of the ultramassive WDs show strong delays on their cooling times.} { Unfortunately, the CO chemical composition in the core of ultramassive WDs still needs to be proven, and the high $^{22}$Ne content that we demand to fit the observations is not consistent with the standard evolutionary channels. We hope that new mechanisms that lead to the formation of ultramassive WDs with higher $^{22}$Ne content can be found in the future.}

\begin{acknowledgements}
This research was partially supported by the MCINN, by the AGENCIA, by the Generalitat de Catalunya, by the STFC and by the CONICET.  ARM acknowledges support from the MINECO under the Ram\'on  y Cajal programme (RYC-2016-20254). ARM and ST acknowledge support from the MINECO AYA2017-86274-P grant, and the AGAUR grant SGR-661/2017. We thank Detlev Koester for providing the atmospheres models to calculate the photometry in Gaia passbands. RR has received funding from the post-doctoral fellowship programme Beatriu de Pin\'os, funded by
the Secretary of Universities and Research (Government of Catalonia) and by the Horizon 2020
programme of research and innovation of the European Union under the Maria Sk\l{}odowska-
Curie grant agreement n. 801370. The research leading to these results has received funding from the European Research Council under the European Union’s Horizon 2020 research and innovation programme n. 677706 (WD3D).

\end{acknowledgements}

\bibliographystyle{aa} 
\bibliography{lowZ}

\end{document}